\documentstyle[12pt]{article}

\newlength{\myleftmargin}
\begin{document}
\renewcommand{\thefootnote}{\fnsymbol{footnote}}
\begin{flushright}
KOBE--FHD--95--06\\
October~~~~~~~~1995
\end{flushright}
\begin{center}
{\large \bf {\boldmath $\Delta s$} density in a proton and unpolarized\\
lepton--polarized proton scatterings}\\
\vspace{3.5em}
T. Morii\footnote[2]{E--mail~~~~morii@kobe--u.ac.jp}\\
\vspace{0.8em}
{\it Faculty of Human Development, Division of}\\
{\it Sciences for Natural Environment}\\
{\it and}\\
{\it Graduate School of Science and Technology,}\\
{\it Kobe University, Nada, Kobe 657, Japan}\\
\vspace{1.5em}
Alexander I. Titov\footnote[3]{E--mail~~~~atitov@thsun1.jinr.dubna.su}\\
\vspace{0.8em}
{\it Bogoliubov Laboratory of Theoretical Physics,}\\
{\it Joint Institute for Nuclear Research,}\\
{\it 141980, Dubna, Moscow region, Russia}\\
\vspace{1em}
and\\
\vspace{1em}
T. Yamanishi\footnote[8]{E--mail~~~~yamanisi@natura.h.kobe--u.ac.jp}\\
\vspace{0.8em}
{\it Research Center for Nuclear Physics,}\\
{\it Osaka University, Ibaraki, Osaka 567, Japan}\\
\vspace{4.5em}
{\bf Abstract}
\end{center}

\baselineskip=24pt

It is shown that the parity--violating deep--inelastic scatterings of
unpolarized charged leptons on polarized protons,
$\ell^{\mp} + \vec P\to \stackrel{\scriptscriptstyle(-)}{\nu_{\ell}} + X$,
could provide a sensitive test for the behavior and magnitude of the polarized
strange--quark density in a proton.
Below charm threshold these processes are also helpful to uniquely determine
the magnitude of individual polarized parton distributions.

\vfill\eject

There have been interesting problems on strange--quark (s--quark) contents in
contemporary hadron physics.
Deductions of the $\sigma$--term from pion-nucleon scatterings imply an
existence of significant s--quark contents in a nucleon\cite{sigmaterm}.
New analysis suggests that about one third of the rest
mass of the proton comes from $s\bar s$ pairs.
So far, interesting experimental proposals\cite{newexp} have been presented to
measure the neutral weak form factors of the nucleon which might be sensitive
to the s--quarks inside the nucleon.
A different idea is also proposed to directly probe the s--quark content of
the proton by using the lepto-- and photo--production of $\phi$--meson that is
essentially 100\% $s\bar s$\cite{phi}.
Another surprizing results on the s--quark contents
in the nucleon have been drawn from the data of polarized deep inelastic
scatterings\cite{PDIS}.
To our surprise, the experimental data have suggested that, contrary to
the prediction of the naive quark model, there is a large and negative
contribution of s--quarks to the proton spin, $i.e.$ $\Delta s=-0.12$, and
furthermore very little of the proton spin is carried by quarks.
For low-energy properties of baryons, conventional phenomenological quark
models treat nucleons as consisting
of only u-- and d--quarks and thus it naturally comes as a big surprise
when some recent measurements and theoretical analyses have indicated
a possible existence of a sizable s--quark.
In order to get deep understanding of hadron dynamics,
it is very important to investigate the behavior of s--quarks in
a nucleon.
In this paper, we concentrate on the behavior of the polarized s--quark
and study the processes sensitive to its polarized distributions
in the nucleon.

So far, several people have suggested various processes, which are sensitive
to polarized s--quark distributions, such as
Drell--Yan processes\cite{Leader}, inclusive $W^{\pm}$-- and
$Z^0$--productions\cite{Soffer94} in polarized proton--polarized proton
collisions, and also inclusive $\pi^{\pm}$-- and $K^{\pm}$--productions in
polarized lepton--polarized proton scatterings\cite{Close}.
However, since the differential cross sections for
Drell--Yan processes and inclusive $W^{\pm}$-- /$Z^0$--hadroproductions are
described by the product of two parton distributions participating in such
processes, one cannot extract the $x$--dependence of polarized s--quark
distributions without ambiguities from such cross sections.
In addition, those of inclusive $\pi^{\pm}$-- and $K^{\pm}$--leptoproductions
include the fragmentation functions of $\pi^{\pm}$-- and $K^{\pm}$--decays
which possess some theoretical ambiguities, and hence it is also difficult to
derive the exact behavior of polarized s--quark distributions from these
processes.
Recently, it has been pointed out that parity--violating polarized electron
elastic scatterings on unpolarized protons can give informations on the
matrix elements, $\langle p|\bar s\Gamma_{\mu}s|p\rangle$ with $\Gamma_{\mu}$
$=\gamma_{\mu}$ and $\gamma_{\mu}\gamma_5$\cite{Fayyazuddin}.
However, since its differential cross section includes not only the
spin--dependent but also spin--independent proton form factors, one
cannot extract the polarized s--quark content without ambiguities
even from such processes.
Here we consider a different process for examining the polarized
s--quark density, which is the parity--violating polarized deep inelastic
scattering at high energy.
It must be advantageous to study such a process because its differential cross
section includes only the spin--dependent structure function of the proton
and is explicitly described as a function of $x$.

In parity--violating deep inelastic scatterings of unpolarized charged
lepton on longitudinally polarized proton, an interesting parameter is the
single--spin asymmetry $A_L^{W^{\mp}}$ defined as
\begin{eqnarray}
A_L^{W^{\mp}}&=&\frac{(d\sigma_{++}^{W^{\mp}}+d\sigma_{-+}^{W^{\mp}})-
(d\sigma_{+-}^{W^{\mp}}+d\sigma_{--}^{W^{\mp}})}
{(d\sigma_{++}^{W^{\mp}}+d\sigma_{-+}^{W^{\mp}})+
(d\sigma_{+-}^{W^{\mp}}+d\sigma_{--}^{W^{\mp}})}\nonumber\\
&=&\frac{d\sigma_{0+}^{W^{\mp}}-d\sigma_{0-}^{W^{\mp}}}
{d\sigma_{0+}^{W^{\mp}}+d\sigma_{0-}^{W^{\mp}}}
=\frac{d\Delta_L\sigma^{W^{\mp}}/dx}{d\sigma^{W^{\mp}}/dx}~,
\label{eqn:A_L}
\end{eqnarray}
where $d\sigma_{0-}^{W^{\mp}}$, for instance, denotes that the lepton is
unpolarized and the helicity of the proton is negative.
Note that since a fast incoming negatively (positively) charged lepton,
$\ell^-$ ($\ell^+$), couples to a $W$--boson only when it has a negative
(positive) helicity, part of spin--dependent cross sections in
eq.(\ref{eqn:A_L}) should be zero.
For parity--violating weak--interacting reactions with $W^{\mp}$ exchanges,
$\ell^{\mp} + \vec P\to \stackrel{\scriptscriptstyle(-)}{\nu_{\ell}} + X$,
the spin--dependent and spin--independent differential cross sections as
a function of momentun fraction $x$ are given by\cite{Anselmino}
\begin{eqnarray}
&&\frac{d\Delta_L\sigma^{W^{\mp}}}{dx}
=16\pi M_NE\frac{\alpha^2}{Q^4}\eta\left\{\pm (\frac{2}{3}+\frac{xM_N}{6E})x~
g_1^{W^{\mp}}(x, Q^2)+(\frac{2}{3}-\frac{xM_N}{12E})~g_3^{W^{\mp}}(x, Q^2)
\right\}~,\nonumber\\
&&\label{eqn:dDs}\\
&&\frac{d\sigma^{W^{\mp}}}{dx}
=16\pi M_NE\frac{\alpha^2}{Q^4}\eta\left\{(\frac{2}{3}-\frac{xM_N}{4E})~
F_2^{W^{\mp}}(x, Q^2)\pm\frac{1}{3}x~F_3^{W^{\mp}}(x, Q^2)\right\}~,
\label{eqn:ds}
\end{eqnarray}
where $E$ is the energy of the charged lepton beam and $M_N$ the mass of the
proton.
$\eta$ is written in terms of the $W$--boson mass $M_W$ as
\begin{equation}
\eta=\frac{1}{2}\left(\frac{G_FM^2_W}{4\pi\alpha}\frac{Q^2}{Q^2+M_W^2}\right)~.
\label{eqn:eta}
\end{equation}
$g_1^{W^{\mp}}$, $g_3^{W^{\mp}}$ in eq.(\ref{eqn:dDs}) and
$F_2^{W^{\mp}}$, $F_3^{W^{\mp}}$ in eq.(\ref{eqn:ds})
represent spin--dependent and spin--independent proton structure functions,
respectively.
Below charm threshold, the region of which could be investigated by SMC
and/or E143 Collabolations, we can describe these structure functions for the
W$^-$ exchange as
\begin{eqnarray}
&&F_2^{W^-}(x, Q^2)=2x\left[c_1\{u_v(x, Q^2)+u_s(x, Q^2)\}+
c_2~\bar d_s(x, Q^2)+c_3~\bar s_s(x, Q^2)\right]~,
\nonumber\\
&&F_3^{W^-}(x, Q^2)=2\left[c_1\{u_v(x, Q^2)+u_s(x, Q^2)\}-
c_2~\bar d_s(x, Q^2)-c_3~\bar s_s(x, Q^2)\right]~,
\label{eqn:str1}\\
&&g_1^{W^-}(x, Q^2)=\left[c_1\{\delta u_v(x, Q^2)+\delta u_s(x, Q^2)\}+
c_2~\delta\bar d_s(x, Q^2)+c_3~\delta\bar s_s(x, Q^2)\right]~,\nonumber\\
&&g_3^{W^-}(x, Q^2)=2x\left[c_1\{\delta u_v(x, Q^2)+\delta u_s(x, Q^2)\}-
c_2~\delta\bar d_s(x, Q^2)-c_3~\delta\bar s_s(x, Q^2)\right]~,\nonumber
\end{eqnarray}
and similarly for the W$^+$ exchange
\begin{eqnarray}
&&F_2^{W^+}(x, Q^2)=2x\left[c_1~\bar u_s(x, Q^2)+c_2\{d_v(x, Q^2)+
d_s(x, Q^2)\}+c_3~s_s(x, Q^2)\right]~,\nonumber\\
&&F_3^{W^+}(x, Q^2)=2\left[-c_1~\bar u_s(x, Q^2)+c_2\{d_v(x, Q^2)+
d_s(x, Q^2)\}+c_3~s_s(x, Q^2)\right]~,
\label{eqn:str2}\\
&&g_1^{W^+}(x, Q^2)=\left[c_1~\delta\bar u_s(x, Q^2)+
c_2\{\delta d_v(x, Q^2)+\delta d_s(x, Q^2)\}+
c_3~\delta s_s(x, Q^2)\right]~,\nonumber\\
&&g_3^{W^+}(x, Q^2)=2x\left[-c_1~\delta\bar u_s(x, Q^2)+
c_2\{\delta d_v(x, Q^2)+\delta d_s(x, Q^2)\}+
c_3~\delta s_s(x, Q^2)\right]\nonumber
\end{eqnarray}
with CKM matrix elements
\begin{equation}
c_1~=~|U_{ud}|^2+|U_{us}|^2~,~~~c_2~=~|U_{ud}|^2~,~~~c_3~=~|U_{us}|^2~.
\label{eqn:kmm}
\end{equation}
Here $\delta q(x, Q^2)=q_+(x, Q^2)-q_-(x, Q^2)$
($q(x, Q^2)=q_+(x, Q^2)+q_-(x, Q^2)$) stands for the polarized (unpolarized)
quark distribution in a proton, and $q_+(x, Q^2)$
($q_-(x, Q^2)$) the quark density having a momentum fraction $x$ with
the helicity parallel (anti--parallel) to the proton helicity.
It should be noticed that since $g_1^{W^-}$ and $g_1^{W^+}$ have no relation to
the flavor singlet axial--vector current, it is not affected by the axial
anomaly which, in deep inelastic reactions with one--photon exchanges, leads
to rather large contributions of the polarized gluons to polarized quark
distribution functions\cite{anomaly}.

In order to examine how the observed parameter is affected by the behavior of
polarized s--quark distributions, we calculate $A_L^{W^{\mp}}$ by substituting
eqs.(\ref{eqn:dDs}) and (\ref{eqn:ds}) into eq.(\ref{eqn:A_L}) as follows,
\begin{equation}
A_L^{W^{\mp}}~=~\frac{\pm (\frac{2}{3}+\frac{xM_N}{6E})x~
g_1^{W^{\mp}}(x, Q^2)+(\frac{2}{3}-\frac{xM_N}{12E})~g_3^{W^{\mp}}(x, Q^2)}
{(\frac{2}{3}-\frac{xM_N}{4E})~F_2^{W^{\mp}}(x, Q^2)\pm\frac{1}{3}x~
F_3^{W^{\mp}}(x, Q^2)}~.
\label{eqn:dDs/ds}
\end{equation}
As typical examples of the polarized s--quark distribution functions, we take
the following three different types;
(i) negative and large $\Delta s$ with
$\Delta s(Q^2=4$GeV$^2)=-0.11$ (BBS model)\cite{BBS}~,
(ii) zero $\Delta s$ with $\Delta s(Q^2=10$GeV$^2$)=0\cite{Cheng90}~,
(iii) positive and small $\Delta s$ with
$\Delta s(Q^2=10$GeV$^2)=0.02$ (KMTY model)\cite{Kobayakawa}~,
where $\Delta s(Q^2)$ is the first moment of $\delta s(x, Q^2)$ and its value
means the amount of the proton spin carried by the s--quark.
All of the models, (i), (ii) and (iii), can reproduce the EMC and SMC data
on $xg_1^p(x)$ equally well.
The $x$--dependence of these distributions are depicted at $Q^2=10$GeV$^2$
in fig.1.
(The explicit forms of $\delta s(x, Q^2)$ are presented in respective
references.)
By using the polarized s--quark distribution of each type, together with the
polarized and unpolarized parton distribution of BBS parametrization\cite{BBS}
for (i), of Cheng--Lai\cite{Cheng90} and DFLM parametrizaton\cite{DFLM} for
(ii), and of KMTY\cite{Kobayakawa} and DO parametrization\cite{Duke} for (iii),
we estimate the $A_{L}^{W^{\mp}}$
at a typical charged lepton energy $E=180$GeV and momentum transfer squared
$Q^2=10$GeV$^2$ whose kinematical region can be covered by SMC experiments.
We see that $A_{L}^{W^-}$ for the $W^-$ exchange process depends on the
behavior of polarized s--quark distributons little because the spin--dependent
proton structure functions, $g_1^{W^-}(x, Q^2)$ and $g_3^{W^-}(x, Q^2)$,
included in $A_{L}^{W^-}$ are dominantly controlled by the polarized valence
u--quark ditribution $\delta u_v(x, Q^2)$ whose magnitude $\Delta u_v$ is much
larger than $\Delta\bar s_s$ in the proton.
However, the situation is quite different for $g_1^{W^+}(x, Q^2)$ and
$g_3^{W^+}(x, Q^2)$ originated from $W^+$ exchanges because $g_1^{W^+}(x, Q^2)$
and $g_3^{W^+}(x, Q^2)$ have no contribution of the polarized valence
u--quark distribution.
Although $g_1^{W^+}(x, Q^2)$ and $g_3^{W^+}(x, Q^2)$ contain the polarized
valence d--quark distribution $\delta d_v(x, Q^2)$, the absolute value of
$\Delta d_v$ is quite smaller than that of $\Delta u_v$.
Thus, $A_{L}^{W^+}$ is expected to be rather sensitive to the
polarized s--quark distribution function compared to the case of $A_{L}^{W^-}$.
The $x$--dependence of $xg_1^{W^+}$, $g_3^{W^+}$ and $A_{L}^{W^+}$ are
calculated and shown in figs.2, 3 and 4, respectively.
From fig.4, one can observe that the behavior of $A_{L}^{W^+}$ significantly
depends on the polarized s--quark distribution for not very small $x$ regions
and hence we can distinguish
the model of polarized s--quark distributions from the data of $A_{L}^{W^+}$.
The reader might consider that the difference obtained here could be
originated from our procedure of having used the different unpolarized
parton distributions for respective models, (i), (ii) and (iii).
In order to examine
if the results are really meaningful, we have carried out the same
calculation using in common the DFLM parametrization for unpolarized
distributions as an example.
Calculated results are presented by lines with small circles in fig.4.
From this result, we can say that the conclusion remains to be unchanged.

However, it should be noted that one cannot directly extract the s--quark
distribution from $xg_1^{W^+}(x, Q^2)$ and $g_3^{W^+}(x, Q^2)$.
This is because, as shown in eq.(\ref{eqn:dDs}), the differential cross
sections are described by a linear combination of $xg_1^{W^+}(x, Q^2)$ and
$g_3^{W^+}(x, Q^2)$ and we cannot measure independently $xg_1^{W^+}(x, Q^2)$
and $g_3^{W^+}(x, Q^2)$ in experiments.
It is interesting to recall that the situation is quite different for the
case of one--photon exchange processes.
Although the differential cross sections are described by the linear
combination of $g_1^p(x)$ and $g_2^p(x)$ in that case as well, $g_2^p(x)$ can
be kinematically neglected compared to $g_1^p(x)$ and hence one can easily
measure $g_1^p(x)$ in experiment.
This is not the case for $W^{\pm}$ exchange processes.
The information on the polarized s--quark distribution reflects to
$A_{L}^{W^+}$ which contains $xg_1^{W^+}(x, Q^2)$ and $g_3^{W^+}(x, Q^2)$.
In practice, the differential cross sections for these
processes are small because of weakly interacting $W$--boson exchanges:
for example, $d\Delta_L\sigma^{W^+}/dx=-0.093$ ($-0.011$, $-0.099$) [pb] and
$d\sigma^{W^+}/dx=0.62$ ($0.75$, $0.52$) [pb] for the type of (iii) ((i),
(ii)) at $x=0.1$, $E=180$GeV and $Q^2=10$GeV$^2$.
Therefore, from the experimental point of view, we must have high luminosity in
order to get informations on the polarized s--quark distribution function
$\delta s(x, Q^2)$ inside a proton.
Futhermore it might be practically rather difficult to identify the missing
events from $\ell^+ + \vec P\to\bar{\nu_{\ell}} + X$.
However, we believe that these difficulties
can be technically overcome.

Another importance for getting $g_1^{W^{\pm}}$ comes from the fact that,
below charm threshold, one can uniquely determine the magnitude of individual
polarized parton density by combining the data on $g_1^{W^{\pm}}$ with the
ones on neutron $\beta$--decay, hyperon $\beta$--decay and the spin--dependent
structure function of proton $g_1^p$.
The following combinations of individual polarized parton content are
well--known,
\begin{eqnarray}
&&\Delta u-\Delta d=a_3~,
\label{eqn:neu}\\
&&\Delta u+\Delta d-2\Delta s=a_8~,
\label{eqn:hyp}\\
&&\frac{4}{18}\Delta u+\frac{1}{18}\Delta d+\frac{1}{18}\Delta s
-\frac{\alpha_s}{6\pi}\Delta g=a_0^{\gamma}~,
\label{eqn:sing}
\end{eqnarray}
where $\Delta g$ in eq.(\ref{eqn:sing}) represents the amount of the proton
spin carried by gluons and is introduced by taking account of
the U$_A(1)$ anomaly of QCD\cite{anomaly}.
Although the values of $a_3$, $a_8$ and $a_0^{\gamma}$ are known from the
experimental data on neutron $\beta$--decay, hyperon $\beta$--decay and the
spin--dependent structure function of proton $g_1^p$, respectively,
it is impossible to determine the magnitude of individual content uniquely
from these equations alone
since there exist four independent variables for three equations.
However, if the values of the first moment of $g_1^{W^+}(x)$ and
$g_1^{W^-}(x)$ can be obtained experimentally
\begin{equation}
\int_0^1 g_1^{W^+}(x)dx+\int_0^1 g_1^{W^-}(x)dx=
c_1~\Delta u+c_2~\Delta d+c_3~\Delta s=a_0^{W^+}+a_0^{W^-}~,
\label{eqn:Wmp}
\end{equation}
then, from four independent equations, (\ref{eqn:neu}),
(\ref{eqn:hyp}), (\ref{eqn:sing}) and (\ref{eqn:Wmp}),
$\Delta u$, $\Delta d$, $\Delta s$ and $\Delta g$ can be determined as follows,
\begin{eqnarray}
&&\Delta u=\frac{(2c_2+c_3)a_3+c_3a_8+2(a_0^{W^+}+a_0^{W^-})}
{2(c_1+c_2+c_3)}~,
\label{eqn:Du}\\
&&\Delta d=\frac{(-2c_1-c_3)a_3+c_3a_8+2(a_0^{W^+}+a_0^{W^-})}
{2(c_1+c_2+c_3)}~,
\label{eqn:Dd}\\
&&\Delta s=\frac{(-c_1+c_2)a_3-(c_1+c_2)a_8+2(a_0^{W^+}+a_0^{W^-})}
{2(c_1+c_2+c_3)}~,
\label{eqn:Ds}\\
&&\Delta g=-\frac{\pi}{3\alpha_s}\frac{1}{2(c_1+c_2+c_3)}\left\{
3(c_1-3c_2-c_3)a_3\right.
\label{eqn:Dg}\\
&&\left. +(c_1+c_2-5c_3)a_8+36(c_1+c_2+c_3)a_0^{\gamma}-
12(a_0^{W^+}+a_0^{W^-})\right\}~.\nonumber
\end{eqnarray}
But how can we actually measure $a_0^{W^{\pm}}$ in experiment?
As long as we remain in the experiment with the longitudinally polarized
proton target, we cannot determine $a_0^{W^{\pm}}$ experimentally.
However, if we consider the cross section with the transversely polarized
proton target, one can obtain $a_0^{W^{\pm}}$ as described in the following.  

The formulas of the transversely polarized cross section have been given by
Anselmino et al.\cite{Anselmino} as follows,
\begin{eqnarray}
\frac{d\Delta_T\sigma^{W^{\pm}}}{dxdyd\phi}&=&
32M_N\frac{\alpha^2}{Q^4}\eta\cos (\psi-\phi)
\sqrt{xyM_N\left\{2(1-y)E-xyM_N\right\}}\nonumber\\
&\times&x(1-y)
\left(\pm g_1^{W^{\pm}}(x, Q^2)+\frac{1}{2x}~g_3^{W^{\pm}}(x, Q^2)\right)~,
\label{eqn:dDs_T}
\end{eqnarray}
where $\phi$ is the azimuthal angle of the lepton in the final state, and
$\psi$ an angle between the proton spin and $x$--axis in the $xy$ plane
orthogonal to the lepton direction ($z$--axis).
These angles must be fixed in principle by suitably arranging experimental
apparatus.
$d\Delta_T\sigma^{W^{\pm}}/dxdyd\phi$ is defined as
\begin{equation}
\frac{d\Delta_T\sigma^{W^{\pm}}}{dxdyd\phi}=
\frac{d\sigma^{W^{\pm}}_{0\uparrow}}{dxdyd\phi}-
\frac{d\sigma^{W^{\pm}}_{0\downarrow}}{dxdyd\phi}~,
\label{eqn:dfntrans}
\end{equation}
where $d\sigma^{W^{\pm}}_{0\uparrow}$ ($d\sigma^{W^{\pm}}_{0\downarrow}$)
denotes that the lepton is unpolarized and the proton is transversely
polarized with its spin orthogonal to the lepton direction at an
angle $\psi$ ($\psi+\pi$) to the $x$--axis.
By integrating eq.(\ref{eqn:dDs_T}) over $y$, one can easily derive the
following formula,
\begin{eqnarray}
\frac{d\Delta_T\sigma^{W^+}}{dxd\phi}-
\frac{d\Delta_T\sigma^{W^-}}{dxd\phi}&=& C~x^{3/2}
\left[\int^1_0 dy\sqrt{y}(1-y)\sqrt{1-y-\frac{xyM_N}{2E}}\right]\nonumber\\
&\times&\left(g_1^{W^+}(x, Q^2)+g_1^{W^-}(x, Q^2)\right)~,
\label{eqn:AT}
\end{eqnarray}
with $C=32\sqrt{2E}M_N^{3/2}\frac{\alpha^2}{Q^4}\eta\cos(\psi-\phi)$.
In eq.(\ref{eqn:AT}), the integral in the square bracket depends on
$x$ alone and can be written by $f(x)$.
Practically, $f(x)$ can be very nicely approximated by the formula,
\begin{equation}
f(x)=0.19635~(1-\frac{2.45}{E}x)~,
\label{eqn:funcf}
\end{equation}
which reproduces the exact result with accuracy better than $10^{-4}$
for $E>50$GeV.
Then, we can get
\begin{equation}
g_1^{W^+}(x)+g_1^{W^-}(x)=\frac{d\Delta_T\sigma^{W^+}/dxd\phi-
d\Delta_T\sigma^{W^-}/dxd\phi}{C~x^{3/2}~f(x)}~,
\label{eqn:trnsasy}
\end{equation}
where the right--hand side of eq.(\ref{eqn:trnsasy}) can be determined
from experiment.
Therefore, if we carry out the experiment with the transversely polarized
target, we can obtain the sum of $a_0^{W^-}$ and $a_0^{W^+}$ by integrating
eq.(\ref{eqn:trnsasy}) over $x$, which leads to the unique determination of
the magnitude
of individual polarized parton densities in a proton.
On the contrary, we can predict the values of $a_0^{W^-}$ and $a_0^{W^+}$
by using each type of
polarized s--quark density.
Some examples are given $a_0^{W^-}=-0.400$ and $a_0^{W^+}=0.798$ for type (i),
$-0.282$ and $0.956$ for type (ii), $-0.196$ and $0.943$ for type (iii) at
$E=180$GeV and $Q^2=10$GeV$^2$.
These predictions should be tested in the forthcoming experiment.

In summary, we have discussed the processes sensitive to the polarized
s--quark distribution and have found that parity--violating reactions with
$W^+$--boson exchange, $\ell^+ + \vec P\to\bar{\nu_{\ell}} + X$, are quite
promising for giving us informations on polarized s--quark distribution
functions inside a proton.
Since the single--spin asymmetry $A_L^{W^+}$ for these processes
significantly depends on the behavior of polarized s--quark distributions,
one can test the behavior and magnitude of
the s--quark polarization by measuring this quantity in experiments.
Futhermore, we have shown that the amount of each quark and gluon carrying
the proton spin can be uniquely determined if the spin--dependent proton
structure functions, $g_1^{W^+}$ and $g_1^{W^-}$, are obtained below charm
threshold by carrying out the experiment with the transversely polarized
target.

Informations on polarized s--quark distributions are decisively important to
understand the proton spin strucutre.
We hope the present predictions could be tested in the forthcoming
experiments.

\vspace{2em}

\begin{center}
{\Large \bf Acknowledgements}
\end{center}

Two of us (T.M. and T.Y.) would like to thank V. V. Burov and the theory
members at JINR, Dubna for their kind hospitality.

\vfill\eject

\begin{center}
{\large \bf Figure captions}
\end{center}
\begin{description}
\item[Fig. 1:] The $x$--dependence of $x\delta s(x, Q^2)$ at
$Q^2=10$GeV$^2$ for various types of polarized s--quark distributions
(i)--(iii). (See text.)
Type (i) is evolved up to $Q^2=10$GeV$^2$.
The dash--dotted, dashed and solid lines indicate the results
calculated for types (i), (ii) and (iii), respectively.

\vspace{2em}

\item[Fig. 2:] The $x$--dependence of spin--dependent proton structure
functions $xg_1^{W^+}$ for partiy--violating reactions with $W^+$--boson
exchanges at $Q^2=10$GeV$^2$ for various types of $\Delta s$.
Various lines represent the same as in fig.1.

\vspace{2em}

\item[Fig. 3:] The spin--dependent proton structure functions $g_3^{W^+}$
as a function of $x$ at $Q^2=10$GeV$^2$.
Various lines represent the same as in fig.1.

\vspace{2em}

\item[Fig. 4:] The single--spin asymmetry $A_L^{W^+}$ as a function of
$x$ at $E=180$GeV and $Q^2=10$GeV$^2$.
The dash--dotted, dashed and solid lines correspond to types (i), (ii)
and (iii), respectively.
The lines added small circles for respective types are the results
calculated by using in common the DFLM parametrization for unpolarized
distributions.
\end{description}

\vspace{2em}

\begin{thebibliography}{1}
\bibitem{sigmaterm}
T. P. Cheng and R. F. Dashen, Phys. Rev. Lett. {\bf 26} (1971) 594;
T. P. Cheng, Phys. Rev. {\bf D13} (1976) 2161;
C. A. Dominguez and P. Langacker, {\it ibid.} {\bf 24} (1981) 190;
J. F. Donoghue and C. R. Nappi, Phys. Lett. {\bf B168} (1986) 105;
J. F. Donoghue, Annu. Rev. Nucl. Part. Sci. {\bf 39} (1989) 1;
J. Gasser, H. Leutwyler, and M. E. Sainio, Phys. Lett. {\bf B253} (1991) 260.
\bibitem{newexp}
D. B. Kaplan and A. Manohar, Nucl. Phys. {\bf B310} (1988) 527;
D. Beck, Phys. Rev. D {\bf 39} (1989) 3248;
R. D. McKeown, Phys. Lett. B {\bf 219} (1989) 140;
Fayyazuddin and Riazuddin, Phys. Rev. {\bf D42} (1990) 794;
D. B. Kaplan, {\it ibid.} {\bf 275} (1992) 137.
\bibitem{phi}
E. M. Henley, G. Krein, and A. G. Williams,
Bull. Am. Phys. Soc. {\bf 34} (1989) 1828;
E. M. Henley, G. Krein, S. J. Pollock and A. G. Williams,
Phys. Lett. {\bf B269} (1991) 31;
E. M. Henley, G. Krein and A. G. Williams,
{\it ibid.} {\bf B281} (1992) 178;
A. I. Titov, Yongseok Oh and Shin Nan Yang, preprint JINR E2--95--48, Dubna,
1995.
\bibitem{PDIS}
J. Ashman et al., EMC Collab. Phys. Lett. {\bf B206} (1988) 364; Nucl. Phys.
{\bf B328} (1989) 1;
B. Adeva et al., SMC Collab., Phys. Lett. {\bf B302} (1993) 533;
P. L. Anthony et al., E142 Collab., Phys. Rev. Lett. {\bf 71} (1993) 959;
D. Adams et al., SMC Collab., Phys. Lett. {\bf B329} (1994) 399;
K. Abe et al., E143 Collab., Phys. Rev. Lett. {\bf 74} (1995) 346.
\bibitem{Leader}
For example,
A. P. Contogouris and S. Papadopoulos, Phys. Lett. {\bf B260} (1991) 204;
E. Leader and K. Sridhar, Phys. Lett. {\bf B311} (1993) 324.
\bibitem{Soffer94}
C. Bourrely and J. Soffer, Nucl. Phys. {\bf B423} (1994) 329.
\bibitem{Close}
F. E. Close and R. G. Milner, Phys. Rev. {\bf D44} (1991) 3691.
\bibitem{Fayyazuddin}
Fayyazuddin and Riazuddin, Phys. Rev. {\bf D42} (1990) 794.
\bibitem{Anselmino}
M. Anselmino, P. Gambino and J. Kalinowski, Z. Phys. {\bf C64} (1994) 267.
\bibitem{anomaly}
G. Altarelli and G. G. Ross, Phys. Lett. {\bf B212} (1988) 391;
R. D. Carlitz, J. C. Collins and A. H. Mueller, Phys. Lett. {\bf B214} (1988)
229; A. V. Efremov and O. V. Teryaev, in {\it Proceedings of the
International Hadron Symposium}, 1988, Bechyn$\check{\rm e}$, Czechoslovakia,
edited by Fischer et al., (Czechoslovakian Academy of Scinece, Prague, 1989).
\bibitem{BBS}
S. J. Brodsky, M. Burkardt and I. Schmidt, Nucl. Phys. {\bf B441} (1995)
197.
\bibitem{Cheng90}
H. Y. Cheng and S. N. Lai, Phys. Rev. {\bf D41} (1990) 91.
\bibitem{Kobayakawa}
K. Kobayakawa, T. Morii, S. Tanaka and T. Yamanishi,
Phys. Rev. {\bf D46} (1992) 2854.
\bibitem{DFLM}
M. Diemoz, F. Ferroni, E. Longo and G. Martinelli, Z. Phys. {\bf C39}
(1988) 21.
\bibitem{Duke}
D. W. Duke and J. F. Owens, Phys. Rev. {\bf D30} (1984) 49.
\end{thebibliography}
\end{document}